\documentclass[11pt]{article}
\usepackage{epsfig}
\usepackage{epstopdf}
\usepackage{amsmath}\usepackage{amssymb}
\usepackage{mathrsfs}
\usepackage{graphicx}

\newcommand {\be}{\begin{equation}}
\newcommand {\ee}{\end{equation}}
\newcommand {\ba}{\begin{eqnarray}}
\newcommand {\ea}{\end{eqnarray}}

\newcommand{\tr}{{\rm tr}}

\newcommand{\GeV}{\mbox{GeV}}

\newcommand{\tz}{{T^0}}

\begin{document}
\def \a'{\alpha'}
\baselineskip 0.65 cm
\begin{flushright}
\ \today
\end{flushright}

\begin{center}{\large
{\bf Constraining Inert Triplet Dark Matter by the LHC and FermiLAT}} {\vskip 0.5 cm} {\bf ${\rm Seyed~ Yaser~ Ayazi}$$^1$ and ${\rm S.~Mahdi~ Firouzabadi}$$^2$}{\vskip 0.5 cm
}
{\small $^1$$School~ of~ Particles~ and~ Accelerators,~ Institute~ for~ Research~ in~
Fundamental$ $ Sciences~(IPM), P.O.~Box~ 19395-5531, Tehran, Iran$
$^2$$Department~ of~ Physics,~ Shahid~ Beheshti~ University,$ $G.~ C.,~ Evin, ~Tehran~ 19839,~ Iran$}

\end{center}

\begin{abstract}
We study collider phenomenology of inert triplet scalar dark matter at the LHC.
We discuss possible decay of Higgs boson to dark matter candidate and apply current experimental data for invisible Higgs decay and $R_{\gamma\gamma}$  to constrain parameter space of our model. We also investigate constraints on dark matter coming from forthcoming measurement, $R_{Z\gamma}$  and mono-Higgs production. We analytically calculate the annihilation cross section of dark matter candidate into $2\gamma$ and $Z\gamma$ and then use FermiLAT data to put constraints on parameter space of Inert Triplet Model. We found that this limit can be stronger than the constraints provided by LUX experiment for low mass DM.

\end{abstract}

\section{Introduction}
The existence of the non-baryonic Dark Matter (DM)  is confirmed by several cosmological observations such as PLANCK, galaxy rotation curves and recent footprints in bullet cluster. However, DM particles have not detected directly yet and the nature of DM still remains unknown. One of the well-known proposals, is to interpret DM as an elementary particle. The SM with minimal Higgs sector can not provide DM candidate. This is a strong motivation in extending SM in a way to provide stable and massive DM particle which is electrically neutral.  The most popular candidate for DM is considered to be the weakly interacting massive particle (WIMP). Various models such as R-parity conserving MSSM, extra dimension models and little higgs model with T-parity provide a WIMP as DM candidate. Commonly, these models are invoked, various particles besides of DM candidate which have not been detected in particle colliders. For the sake of  simplicity, SM extensions with singlet scalar or fermion fields are strongly motivated. In the WIMP scenario, DM candidate can produce required relic density which is measured by PLANCK satellite\cite{PLANCK}. In Ref~\cite{SDM} and \cite{Kim:2006af}, it is shown that allowed region for parameters space of singlet scalar and fermionic DM are strictly limited by relic density constraints. The next simplest candidate for DM is $SU(2)_L$ scalar triplet field. In this model, the lightest component of triplet field is neutral and provides suitable candidate for DM. In \cite{ITM}, it has been shown, for DM mass lower than $7~\rm TeV$, relic abundance agree with PLANCK data.

Aside from PLANCK data, direct searches for DM  at underground experiments and negative results in searches for new physics reported by CMS and ATLAS experiments at the LHC are crucial in interpretation of DM results. Since the Higgs boson can participate in DM-nucleon scattering and DM annihilation, the discovery of Higgs boson and measurements of its decay rates would set limit on any beyond SM that provides a DM candidate.

In this paper, we extend SM by a $SU(2)_L$ triplet scalar with hypercharge $Y=0,2$ and consider its lightest component as DM particle.
 Study of the relic density as well as the direct detection of the dark matter in this model has been done in \cite{ITM}. Proceeding their work, we focus on  parameters space is allowed by PLANCK data and study new constraints which obtain from collider experiments, direct and indirect detection.

This paper is structured as follows: In the next section, we introduce the model and review direct detection constraints on it. In
section~3, we study collider phenomenology of inert triplet DM and constraints which  arise from  experimental observables at LEP and LHC.
Then we calculate annihilation of inert triplet scalar DM into $\gamma\gamma$ and $Z\gamma$ in our galaxy center and apply recent result of FermiLAT in constraining our model. The conclusions are given
in section~4. The form factor formulae for calculating the decay rate of Higgs boson to $2\gamma$ and $Z\gamma$ are summarized in the appendix.

\section{Inert Triplet Model}
The Inert Triplet model (ITM) is an extension of the SM that can provide DM particle. In this model, apart from the SM Higgs doublet, we add a $SU(2)_L$ triplet scalar with $Y=0$ or $Y=2$. In addition we impose $Z_2$ symmetry condition under which the triplet is odd and all the SM fields  are even.  The $Z_2$ symmetry is not spontaneously broken since the triplet does not develop a vacuum expectation value. The triplet for $Y=0$ can be parameterized as:
\begin{eqnarray}
T=\left(\begin{array}{cc}
   \frac{1}{\sqrt{2}}T^0 & -T^{+} \\
   -T^- & -\frac{1}{\sqrt{2}}\tz
  \end{array}\right),
\end{eqnarray}
 where  $\langle T^0\rangle=0$. The relevant scalar potential which is allowed by $Z_2$ symmetry can be written as:
\begin{eqnarray}
V&=&m^2 |H|^2 + M^2 \tr[T^2] + \lambda_1 |H|^4\nonumber+ \lambda_2 (\tr[T^2])^2+ \lambda_3 |H|^2 \tr[T^2] \label{potentioal Y=0}.
\end{eqnarray}
The vacuum expectation value in SM sector of ITM is given by:
\begin{eqnarray}
\langle H\rangle=\frac{1}{\sqrt{2}} \left(\begin{array}{cc}
   0 \\
  v
  \end{array}\right),~~\ v=246 ~\GeV.
\end{eqnarray}
In order for the vacuum to be bounded from below, we demand the following conditions on the parameters:
\begin{eqnarray}
 \lambda_1,\ \lambda_2 \geq 0 ,~~\,(\lambda_1 \lambda_2)^{1/2}-\frac{1}{2}|\lambda_3|>0\ .
\end{eqnarray}
The conditions for local minimum are satisfied for $m^2 <0$, $v^2 = - m^2/2\lambda_1$ and $2M^2 + \lambda_3 v^2 > 0$. The triplet masses can be written by two parameters $\lambda_3$ and $M$:
\begin{eqnarray}
m_{T^0}^2 = m_{T^{\pm}}^2 = M^2 + \frac{1}{2}\lambda_3 v^2.\label{mDM}
\end{eqnarray}

At tree level, as it is seen in the above relation, all the components of $T$ own the same mass, but at loop level the charged components are slightly heavier than $T^0$. In \cite{Minimal dark matter}, it was shown that the mass splitting between charged and neutral components in  the case of zero hypercharge is:
\begin{eqnarray}
\Delta m=(166\pm1) \rm~ MeV. \label{eq:gap}
\end{eqnarray}
Note that the $Z_2$ symmetry ensures the stability of the lightest component to act as a cold DM candidate.
The scalar and gauge interactions of ITM have been extracted in terms of real fields in \cite{ITM}.

In case $Y=2$ the $SU(2)_L$ triplet can be parameterized as:
\begin{eqnarray}
T=\left(\begin{array}{cc}
   \frac{1}{\sqrt{2}}T^+ & T^{++} \\
  T^0_r + iT^0_i & -\frac{1}{\sqrt{2}}T^+
  \end{array}\right)
\end{eqnarray}
and the $Z_2$ invariant form of potential is:
\begin{eqnarray}
V&=&m^2 |H|^2 + M^2 \tr[T^\dag T] + \lambda_1 |H|^4 + \lambda_2 \tr[T^\dag T T^\dag T] + \lambda_3 \left( \tr[T^\dag T] \right)^2\nonumber\\
 &+&\lambda_4 |H|^2\ \tr[T^\dag T] + \lambda_5 H^\dag T T^\dag H.
\label{Y=2}
\end{eqnarray}
\subsection{Direct Detection}
In case of $Y=0$, the $T^0$ DM can interact with nucleon by exchanging Higgs boson. The spin independent cross section of  DM-nucleon is given by \cite{DM lattice QCD}:
\begin{equation}
\sigma_{SI} = \frac{\lambda^2_3f_n^2m^2_N}{4\pi}\frac{\mu^2m^2_N}{m^2_{T^0}m^4_h},
\label{si}
\end{equation}
where the coupling constant $f_n$ is given by nuclear matrix elements\cite{Higgs-nucleon} and $\mu=m_Nm_{T^0}/(m_N+m_{T^0})$ is the reduced mass of DM-nucleon.
There are several experiments to detect DM particles directly through the elastic DM-nucleon scattering. The strict bounds on the DM-nucleon cross section obtained from $\rm XENON100$\cite{XENON100} and $\rm LUX$\cite{LUX} experiments. The minimum upper limits on the spin independent cross section are:
\begin{eqnarray}
{\rm XENON100}:  \sigma_{SI}&\leq& 2\times 10^{-45} \rm cm^2
\nonumber\\ {\rm LUX}: \sigma_{SI}&\leq& 7.6\times 10^{-46} \rm cm^2.
\label{Direct}
\end{eqnarray}

In $Y=2$ case, the $T^0_r$ or $T^0_i$ are playing the role of DM particle. Due to gauge coupling of $Z$ to $T^0_{r,i}$, the DM-nucleon cross section is $10^{-38}~\rm cm^2$ and much larger than upper limits by XENON100 experiment. This excludes all the regions of parameter space except $m_{DM}<1 ~\rm GeV$.

Fig.~\ref{scater}, depicts allowed region in DM mass and $\lambda_3$ couplings plane which does not violate $90\%$ C.L experimental upper bounds of $\rm XENON100$ and $\rm LUX$ for $m_Z/2<m_{T^0}<m_h/2$ and $m_{T^0}>m_h/2$. In this figure, we compare these bounds with other constraints which arise from LHC observables.
\section{Collider Phenomenology of Inert Triplet Dark Matter}
In this section, we will study the constraints coming from  experimental observables at LEP and LHC. The triplet scalar can contribute to several observables at colliders. As it was mentioned, in context of ITM, in mass regimes lower than $7~\rm TeV$, relic density conditions are satisfied. Since direct detection constraints are weak for large masses ($m_{DM}>~1\rm TeV$) henceforward, we assume $m_{DM}<1~\rm TeV$ and evaluate other experimental constraints on parameters space for low mass DM.

A common approach to study beyond SM is considering the electroweak precision test. It is shown that contribution of ITM on oblique parameters S and T is negligibly small\cite{ITM}. In the following, we study constraints that arise from other observables.

The most constraining observable for ITM parameters is the Z boson decay width. The Z boson decay width was measured to be $\Gamma_Z=2.4952\pm0.0023~\rm GeV$, is in good agreement with the SM prediction. This will constrain the contribution of $T^{\mp}T^{\pm}$ to Z width. The decay rate of $Z\rightarrow T^{\mp} T^{\pm}$ is given by:
\begin{eqnarray}
\Gamma(Z\rightarrow T^{\mp} T^{\pm}))& =\frac{g^2c^2_W m_Z}{\pi}(1-\frac{4m^2_{T^{\pm}}}{m^2_Z})^{3/2},
\label{amp1}
\end{eqnarray}
where $g$ is the weak coupling and $c_W=\cos\theta_W$. Since $Z\rightarrow T^{\mp}T^{\pm}$ is suppressed for $m_{T^{\pm}}<m_Z/2$, we assume that $m_{T^0}, m_{T^{\pm}}>45.5~\rm GeV$.

\subsection{Invisible Higgs decays}
\begin{figure}
\begin{center}
\centerline{\hspace{2.5cm}\epsfig{figure=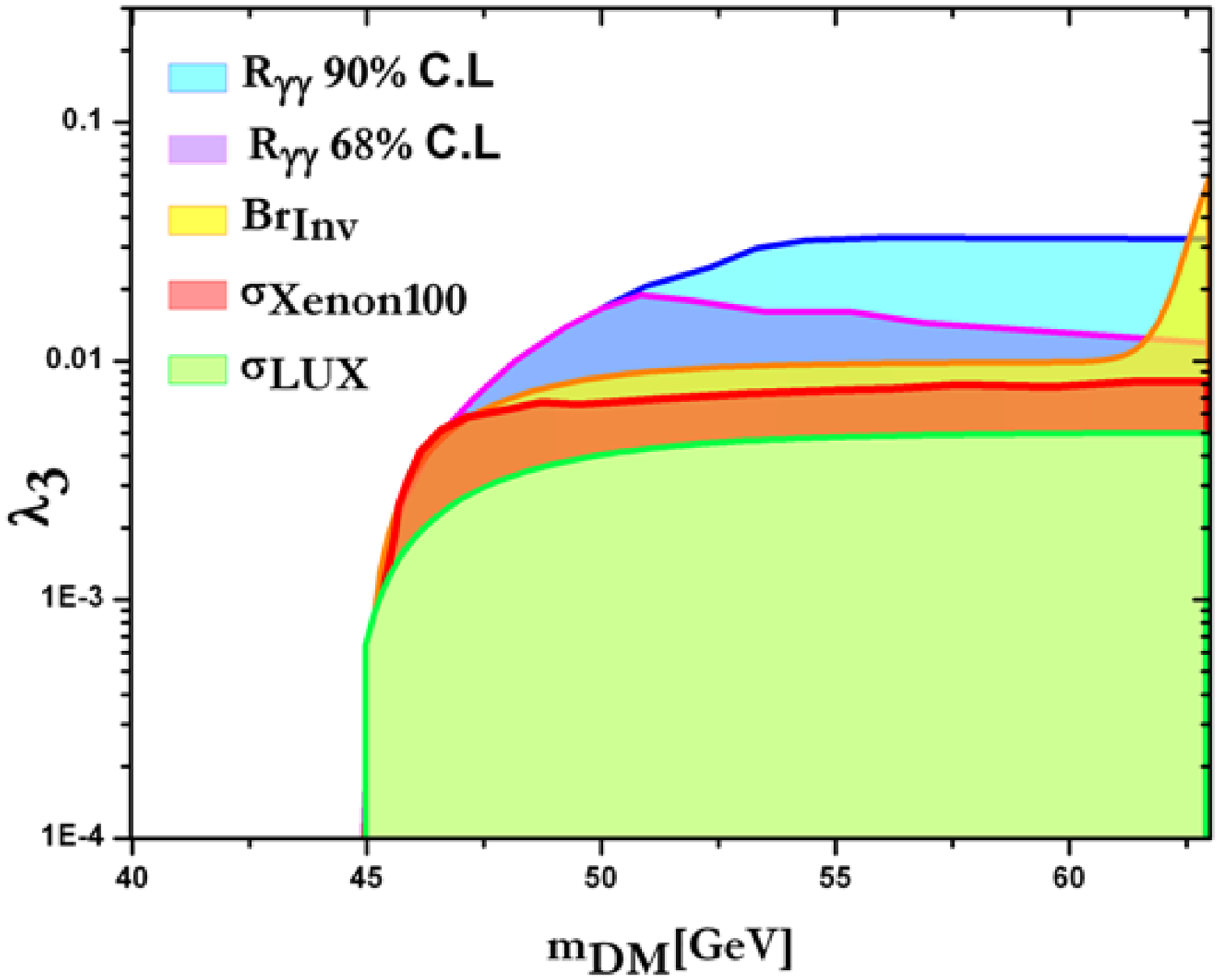,width=8.5cm}\hspace{-2.5cm}\epsfig{figure=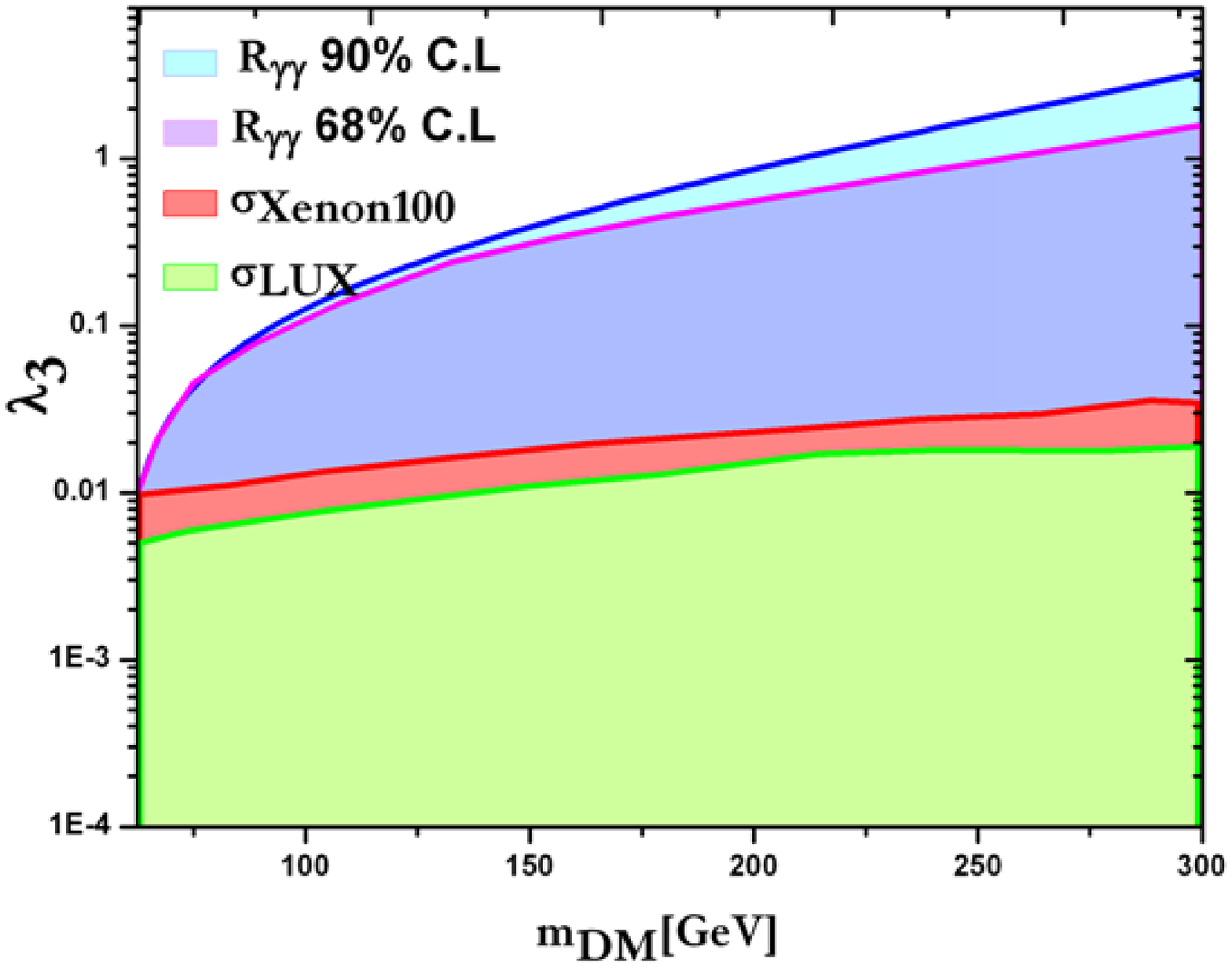,width=8.5cm}}
\centerline{\vspace{-1.5cm}\hspace{0.5cm}(a)\hspace{6cm}(b)}
\centerline{\vspace{-0.0cm}}
\end{center}
\caption{Shaded areas depict ranges of parameter space in mass of DM and $\lambda_3$ coupling plane which are consistent with
experimental measurements of $R_{\gamma\gamma}$ with $90\%$ and $68\%$ C.L, upper limit on $Br(h\rightarrow \rm Invisible)$ with $95\%$ C.L, upper limit on $\sigma_{\rm Xenon100}$ and $\sigma_{\rm LUX}$ with $90\%$ C.L. a) for $45.1<m_{DM}<62.5$, b) for $62.5<m_{DM}$.}\label{scater}
\end{figure}

Invisible Higgs decays provide chance for exploring possible DM-Higgs boson coupling. Nevertheless, invisible Higgs boson decays are not sensitive to DM coupling when $m_{T^0}>m_h/2$. Any components of triplet scalar lighter than SM higgs boson can contribute to the invisible decay mode of higgs boson. The branching ratio is:
\begin{eqnarray}
Br(h\rightarrow \rm Invisible)& =\frac{\Gamma(h\rightarrow \rm Inv)_{\rm SM}+\Gamma(h\rightarrow 2T^0)}{\Gamma(h)_{\rm ITM}},
\label{decayinv1}
\end{eqnarray}
where we have used the following expression for $\Gamma(h)_{\rm ITM}$ (total decay width of higgs boson in ITM):
\begin{eqnarray}
\Gamma(h)_{\rm ITM}=\Gamma(h)_{\rm SM}+\sum_{\chi=T^0,T^{\pm},\gamma}\Gamma(h\rightarrow 2\chi).
\label{Total}
\end{eqnarray}
Total width of higgs boson in SM is $\Gamma(h)_{\rm SM}=4.15 ~ \rm MeV$ \cite{SM Higgs branching ratio} and the partial width for $h\rightarrow 2T^0$ is given by:
\begin{eqnarray}
\Gamma(h\rightarrow 2T^0)& =\frac{\lambda^2_3v^2_0}{4\pi m_h}\sqrt{1-\frac{4m^2_{T^0}}{m^2_h}}.
\label{decayinv1}
\end{eqnarray}
The partial width for $h\rightarrow T^{\mp}T^{\pm}$ and $h\rightarrow 2\gamma$ will be given in Eqe.~\ref{gamma} and \ref{decayinv2}.
The SM prediction for branching ratio of the Higgs boson decaying to invisible particles caused by $h\rightarrow ZZ^*\rightarrow 4\nu$  is \cite{Higgs branching ratio}:
\begin{eqnarray}
Br(h\rightarrow ZZ^*\rightarrow 4\nu)=1.2\times10^{-3}.
\label{amp1}
\end{eqnarray}
Recently,  ATLAS Collaboration has performed a search of the SM higgs boson in its invisible decay mode and obtained  an upper limit of $75\%$ with $95\%$
C.L, at $125.5 \rm GeV$ for $Br(h\rightarrow \rm Invisible)$ \cite{Invisible Higgs decay Exp}. Since invisible higgs decay is forbidden kinematically for $m_{D}>m_{h/2}$, we study  $Br(h\rightarrow \rm Invisible)$ separately in Fig.~\ref{scater}-a. In this figure, we suppose $m_Z/2<m_{T^0}<m_h/2$ and show  valid area in mass of DM and $\lambda_3$ coupling plane which is consistent with  experimental upper limit on $Br(h\rightarrow \rm Invisible)$(with $95\%$ C.L).
As it is seen, allowed region for $m_{DM}\rightarrow m_h/2$ extends because of Eq.~\ref{decayinv1} $\lambda^2_3\varpropto1/(1-\frac{4m^2_{T_0}}{m^2_h})^{1/2}$. Note that according to Eq.~\ref{mDM}, $\lambda^2_3$ can not reach infinity for $m_{DM}\rightarrow m_h/2$. It is notable that allowed region of $Br(h\rightarrow \rm Invisible)$ and direct detection experiments are very similar for $m_Z/2<m_{T^0}<m_h/2$.

\subsection{$R_{\gamma\gamma}$ constraints on ITM dark matter}
ITM also contribute to partial width of $\Gamma_{h\rightarrow\gamma\gamma}$. The partial decay rate for this process has not been measured at the LHC but the ratio of the diphoton rate of the observed Higgs to the SM prediction have been measured recently by CMS\cite{CMS R} and ATLAS\cite{ATLAS R} collaborations:
\begin{eqnarray}
{\rm CMS}:  R_{\gamma\gamma}&=& \frac{\sigma_{\rm measured}}{\sigma_{SM}}=1.14 ^{+0.26}_{-0.23},
\nonumber\\ {\rm ATLAS}: R_{\gamma\gamma}&=&\frac{\sigma_{\rm measured}}{\sigma_{SM}}=1.17\pm0.27.
\label{R}
\end{eqnarray}
The diphoton cross section normalized to SM prediction has been defined in the ITM as follow:
\begin{equation}
R_{\gamma\gamma} = \frac{\sigma(pp\rightarrow h\rightarrow \gamma\gamma)_{\rm ITM}}{\sigma(pp\rightarrow h\rightarrow \gamma\gamma)_{\rm SM}}\simeq\frac{\Gamma(h\rightarrow \gamma\gamma)_{\rm ITM}\times\Gamma(h)_{\rm SM}}{\Gamma(h\rightarrow \gamma\gamma)_{\rm SM}\times\Gamma(h)_{\rm ITM}},
\label{si}
\end{equation}
where $\Gamma(h\rightarrow \gamma\gamma)_{\rm SM}$ and $\Gamma(h\rightarrow \gamma\gamma)_{\rm ITM}$ are partial decay rate of process $h\rightarrow \gamma\gamma$ in context of SM and ITM respectively. Note that the largest contribution to the production of Higgs boson at the LHC is through gluon fusion. In above definition, we have used the fact that cross section of Higgs production in ITM is similar to SM. In ITM, for $m_Z/2<m_{T^0}<m_h/2$ there are two sources of deviation from $R_{\gamma\gamma}=1$. First is partial decay rate $h\rightarrow \gamma\gamma$ caused by charged scalar $T^{\pm}$ in loop level. The Feynman diagrams for Higgs decay to $\gamma \gamma$ depict in Fig.~\ref{decay}. The decay rate is determined by:
\begin{eqnarray}
\Gamma(h\rightarrow \gamma\gamma)& =\frac{G_f\alpha^2m^3_h}{128\sqrt{2}\pi^3}|\frac{4}{3}{\cal{A}}_{1/2}(x_i)+{\cal{A}}_{1}(x_i)+2v_0\lambda_3\frac{ g m_W}{c^2_Wm^2_{T^{\pm}}}{\cal{A}}_{0}(x_i)|^2,
\label{gamma}
\end{eqnarray}
where $G_f$, is the Fermi constant. The formulae for form factors ${\cal{A}}_{i}(x_j)$  are given in Eqs.~\ref{function1} of the appendix. The form factors ${\cal{A}}_{1/2}$, ${\cal{A}}_{1}$  induced by top quark and $W$ gauge boson loops (SM contributions) \cite{gammagamma} and ${\cal{A}}_{0}(x_i)$ arises from $T^{\pm}$ loop. Notice that the interference between SM and ITM contributions can be either constructive or destructive which lead to a decrease or
an increase of the  $\Gamma (h\rightarrow \gamma\gamma)$.

Other sources of deviation from  $R{\gamma\gamma}=1$ are possible decay $h\rightarrow T^0T^0$ and $h\rightarrow T^{\pm}T^{\pm}$ which contribute to total decay rate of Higgs boson in ITM. The $\Gamma(h\rightarrow T^0T^0)$ is given in Eqe.~\ref{decayinv1} and $\Gamma(h\rightarrow T^{\pm}T^{\pm})$ is expressed as:
\begin{eqnarray}
\Gamma(h\rightarrow T^{\pm}T^{\pm})& =\frac{\lambda^2_3v^2_0}{\pi m_h}\sqrt{1-\frac{4m^2_{T^0}}{m^2_h}}.
\label{decayinv2}
\end{eqnarray}

Direct detection  and invisible higgs decay measurements have shown their data to be consistent with the background-only
hypothesis. This  allows to set a $90\%$ confidence limits upper limit on the cross section WIMP-nucleon elastic scattering and $95\%~\rm C.L$ for higgs decay. The values of CMS and ATLAS $R_{\gamma\gamma}$ measurements are in one $\sigma$ confidence interval ($68\%~\rm C.L$). To compare result of direct detection experiment to $R_{\gamma\gamma}$ results, We perform a $\chi^2$-fit analysis for $R_{\gamma\gamma}$ as a function of DM mass and  $\lambda_3$ coupling. This quantity has been defined as:
\begin{eqnarray}
\chi^2=(\frac{R_{\gamma\gamma}^{ITM}-R_{\gamma\gamma}^{exp}}{\sigma_{exp}})^2,
\label{Xai}
\end{eqnarray}
where $R_{\gamma\gamma}^{ITM}$ is ITM prediction for $R_{\gamma\gamma}$, $R_{\gamma\gamma}^{exp}$ is CMS measurements and $\sigma_{exp}$ is total experimental uncertainty which reported by CMS measurements \cite{CMS R}. In Fig.~\ref{scater}-a, we selected points in mass of DM and $\lambda_3$ coupling plane for which are consistent with CMS experimental measurement of  $R_{\gamma\gamma}$ with $90\%$ and $68\%$ confidence interval. We consider CMS measurement because absolute uncertainty of this experiment is smaller than ATLAS. As it is seen, for $m_Z/2<m_{T^0}<m_h/2$ allowed region is not much different from other measurements. Since decay rate for $h\rightarrow \gamma\gamma$ is one order of magnitude smaller than  $h\rightarrow 2T^0$, the main contribution to $R_{\gamma\gamma}$ comes from $\Gamma (h)_{\rm ITM}$ which appears in denominator. For $m_{T^0}>m_h/2$, $h\rightarrow 2T^0$ is forbidden, and $R_{\gamma\gamma}$ only depend on the $\Gamma(h\rightarrow \gamma\gamma)$. Fig.~\ref{scater}-b demonstrate allowed region in mass of $T^0$  and $\lambda_3$ plane for $m_{T^0}>m_h/2$.

\subsection{$R_{\gamma Z}$ constraints on ITM dark matter}
We pursue our analysis on ITM phenomenology by calculating the $h\rightarrow Z\gamma$ decay.
Similar to the case of $R_{\gamma\gamma}$, ITM can affect on $R_{\gamma Z}$. We define normalized ratio of Z-photon cross section $R_{\gamma Z}$ as follow:
\begin{equation}
R_{Z\gamma} = \frac{\sigma(pp\rightarrow h\rightarrow Z\gamma)_{\rm ITM}}{\sigma(pp\rightarrow h\rightarrow Z\gamma)_{\rm SM}}=\frac{\sigma (gg\rightarrow h)_{\rm ITM}\times Br(h\rightarrow Z\gamma)_{\rm ITM}}{\sigma (gg\rightarrow h)_{\rm SM}\times Br(h\rightarrow Z\gamma)_{\rm SM}},
\label{Zgamma}
\end{equation}
where $\sigma(pp\rightarrow h\rightarrow Z\gamma)_{\rm ITM}$ is total cross section. For ITM similar to SM, the main production channel is gluon fusion.

\begin{figure}
\begin{center}
\centerline{\hspace{1cm}\epsfig{figure=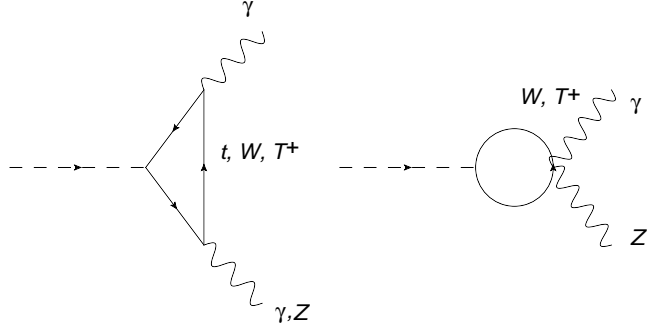,width=8.5cm}}
\centerline{\vspace{-0.8cm}}
\end{center}
\caption{The Feynman diagrams for Higgs decay to $\gamma \gamma$ and $\gamma Z$.}\label{decay}
\end{figure}

Feynman diagrams for Higgs decay to $\gamma Z$ depict in Fig.~\ref{decay}. The decay rate for $h\rightarrow Z\gamma$ can be expressed by:
\begin{eqnarray}
\Gamma(h\rightarrow Z\gamma)&=&\frac{G_f\alpha^2 M^3_h}{16\sqrt{2}\pi^3}(1-\frac{M^2_Z}{M^2_h})^3|{\cal{A}}_{t}(x_i,y_i)+{\cal{A}}_{W}(x_i,y_i)\nonumber \\&+&{\cal{A}}_{T^+}(x_i,y_i)|^2,
\label{amp1}
\end{eqnarray}
where the formulae for form factors ${\cal{A}}_{t}(x_i,y_i)$, ${\cal{A}}_{W}(x_i,y_i)$ and ${\cal{A}}_{T^+}(x_i,y_i)$  are given in Eqs.~\ref{formfactor} of the appendix. The form factors ${\cal{A}}_{t}$ and ${\cal{A}}_{W}$ for the first time have been calculated in context of SM in\cite{gammaZ}. We have calculated the form factor ${\cal{A}}_{T^+}$ which is shown in Fig.~\ref{decay}.

\begin{figure}
\begin{center}
\centerline{\hspace{-0.5cm}\epsfig{figure=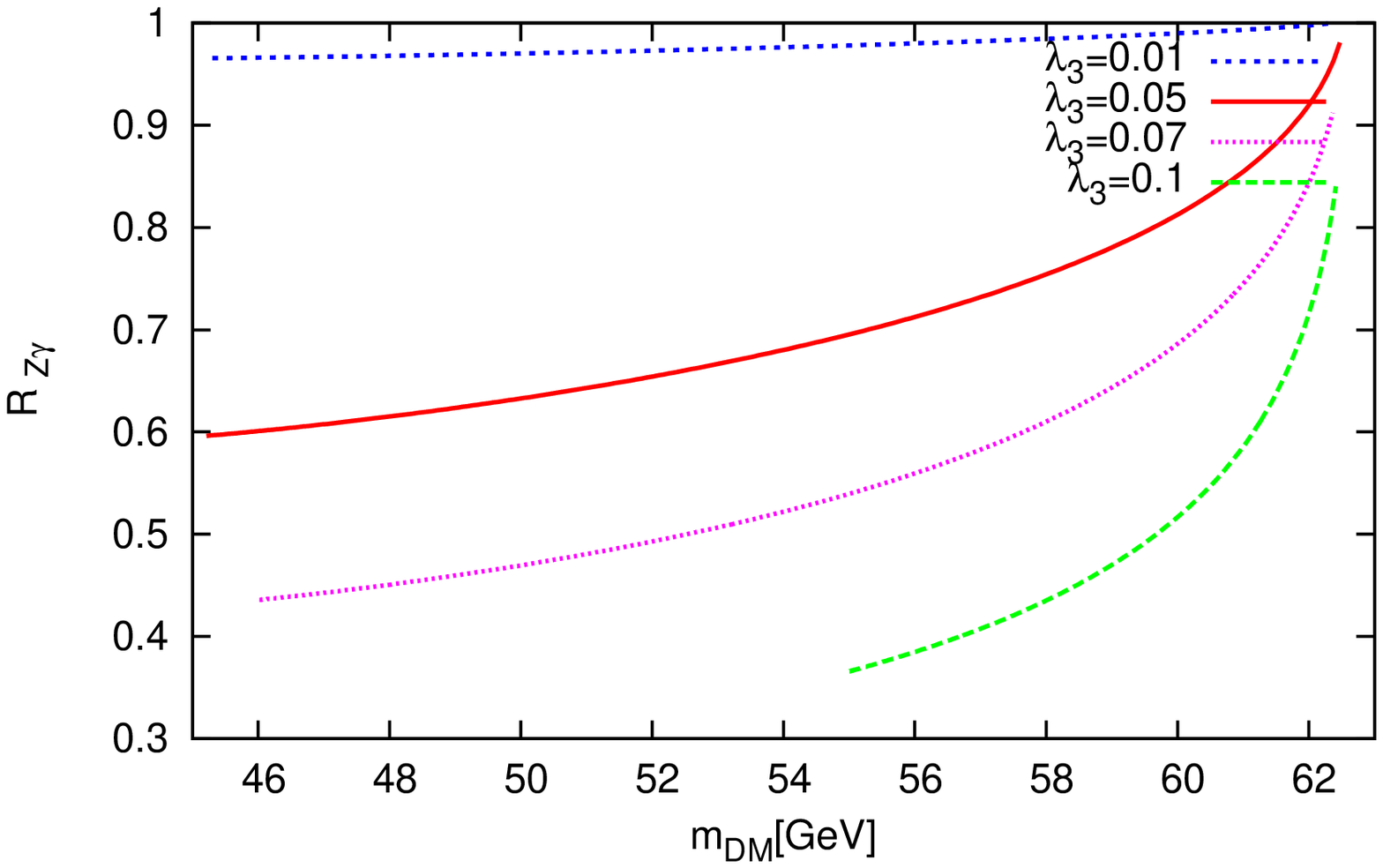,width=6.5cm}\hspace{0cm}\epsfig{figure=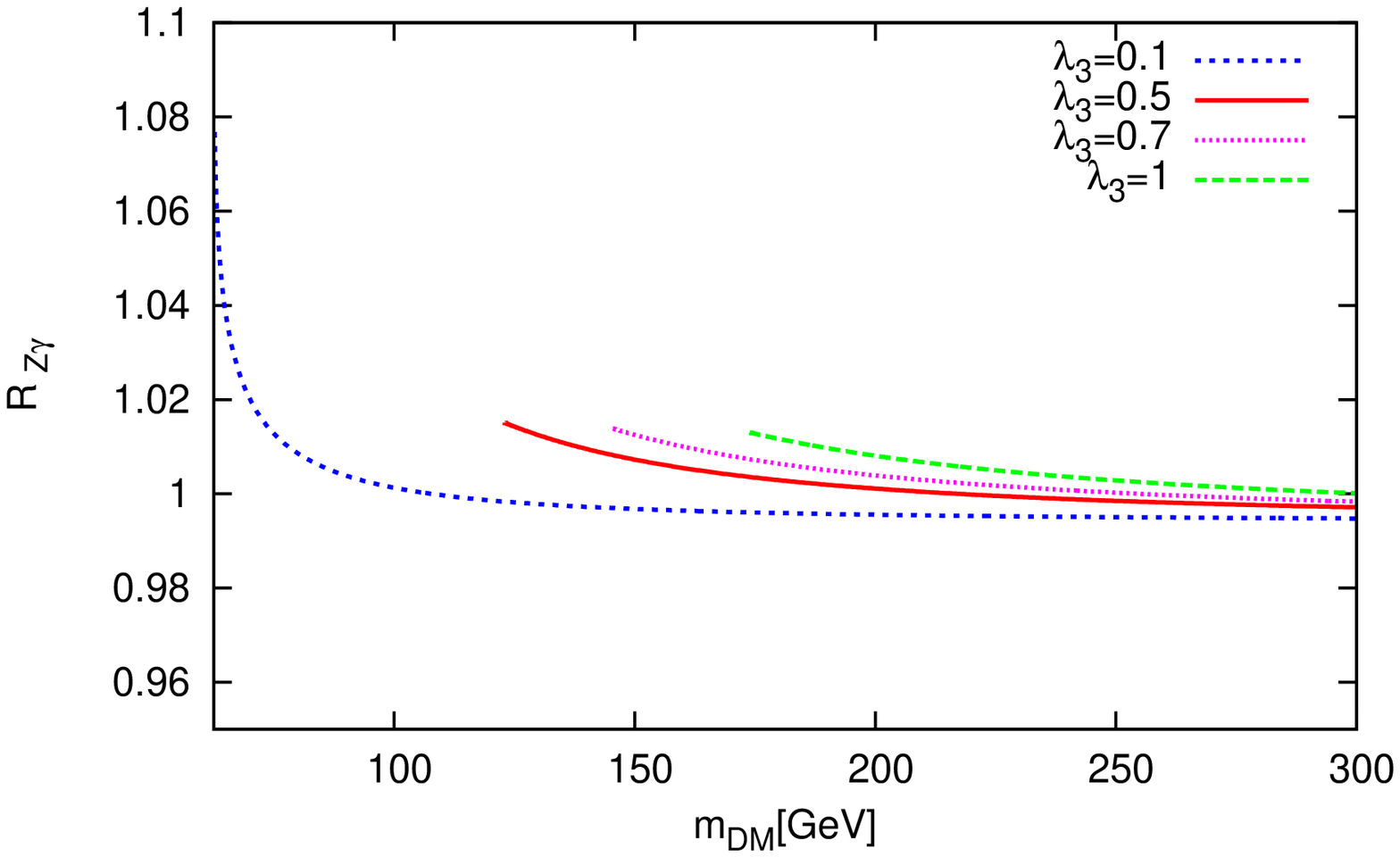,width=6.5cm}}
\centerline{\vspace{-2cm}\hspace{0.5cm}(a)\hspace{6cm}(b)}
\centerline{\vspace{0.8cm}}
\end{center}
\caption{The $R_{\gamma Z}$ as a function of
the DM mass for several values of $\lambda_3$.  a) for $45.1<m_{DM}<62.5$, b) for $62.5<m_{DM}$.}\label{RZgamma}
\end{figure}

 ATLAS and CMS collaborations have presented a search for the SM Higgs boson in the decay channel $h\rightarrow Z\gamma$\cite{RgammaZ}. Sensitivity of these  measurements are far from SM prediction. For a Higgs boson mass of $125~\rm GeV$, the observed exclusion limits are between $7.3$ and $22$ times of the SM prediction\cite{RgammaZ}. Nevertheless, since it is sensitive to mass of charged scalar in ITM, the $h\rightarrow Z\gamma$ decay channel is worth to analyze. In Fig.~\ref{RZgamma}, we depict dependency of  $R_{\gamma Z}$ versus the DM mass for several values of $\lambda_3$. As it is seen, for  $45.1<m_{DM}<62.5$, due to dependency of $R_{\gamma Z}$ to $h\rightarrow 2T^0$, destructive effect of ITM to $R_{\gamma Z}$ can be large. As a result, if forthcoming observed exclusion limit is in the range of SM prediction, ITM parameters can be constrained by this measurement. For $62.5<m_{DM}$, the only contribution to $R_{\gamma Z}$ comes from $h\rightarrow Z\gamma$ decay and its effect is constructive.

\subsection{Mono-X searches at the LHC}
Since DM particles escape the detectors, DM production signature at colliders appears as missing transverse energy ($E_T$). Recently several mono-X searches at the LHC have studied different $X+E_T$ signals, which $X$ can be jet \cite{mono-jet}, Higgs boson \cite{mono-higgs}, $Z/W$ boson \cite{mono-W} or photon \cite{mono-photon}.

In this section, we explore the LHC phenomenology of $T^0$ DM production accompanied with X. In ITM with $Y=0$, $T^0$ do not interact to Z, photon, gluon and quark at the Lagrangian level. Also coupling of $T^0$ to $W$ boson only depends on weak coupling and therefore mono-W production can not constrain $\lambda_3$. Also as it is mentioned in previous section, invisible Higgs decay is  the best process for exploring DM and Higgs coupling but this channel is not sensitive to mass of DM for $ m_{DM}>m_h/2$. Therefore it is worthwhile to study other Higgs boson observables such as mono-Higgs production.

Recently in \cite{mono-higgs}, it has been performed a background study for mono-Higgs signal at the LHC  shows that the LHC can be sensitive to various beyond SM. The amplitude for Mono-Higgs production with two $T^0$ particles is given by:
\begin{eqnarray}
{\cal{|M|}}^2_{Mono-higgs} & =\frac{m^2_q\lambda^2_3}{2v^2_0(4S-m^2_h)}[2S-m^2_q],
\label{amp1}
\end{eqnarray}
where $\sqrt{S}$ is partonic center-of-mass energy. The total cross section of Mono-Higgs production involving DM particles at hadron colliders
can be computed by considering the partonic cross section with
the parton distribution functions (PDF) for the initial hadrons.
To obtain $\sigma(pp\rightarrow hT^0T^0)$, we have employed
the MSTW parton structure functions \cite{MSTW}. The total cross section
for production of $hT^0T^0$ at the $pp$ collision is expressed by:
\begin{eqnarray}
\sigma(pp\rightarrow hT^0T^0)=\sum_{ab} \int
dx_1dx_2f_a(x_1,Q^2)f_b(x_2,Q^2) \widehat{\sigma}(ab\rightarrow
hT^0T^0), \
\end{eqnarray}
where $x_i$ are the parton momentum fractions, $Q$ is the factorization scale and $f_{a,b}(x_i,Q^2)$ are the PDFs of proton.

\begin{figure}
\begin{center}
\centerline{\hspace{-0.5cm}\epsfig{figure=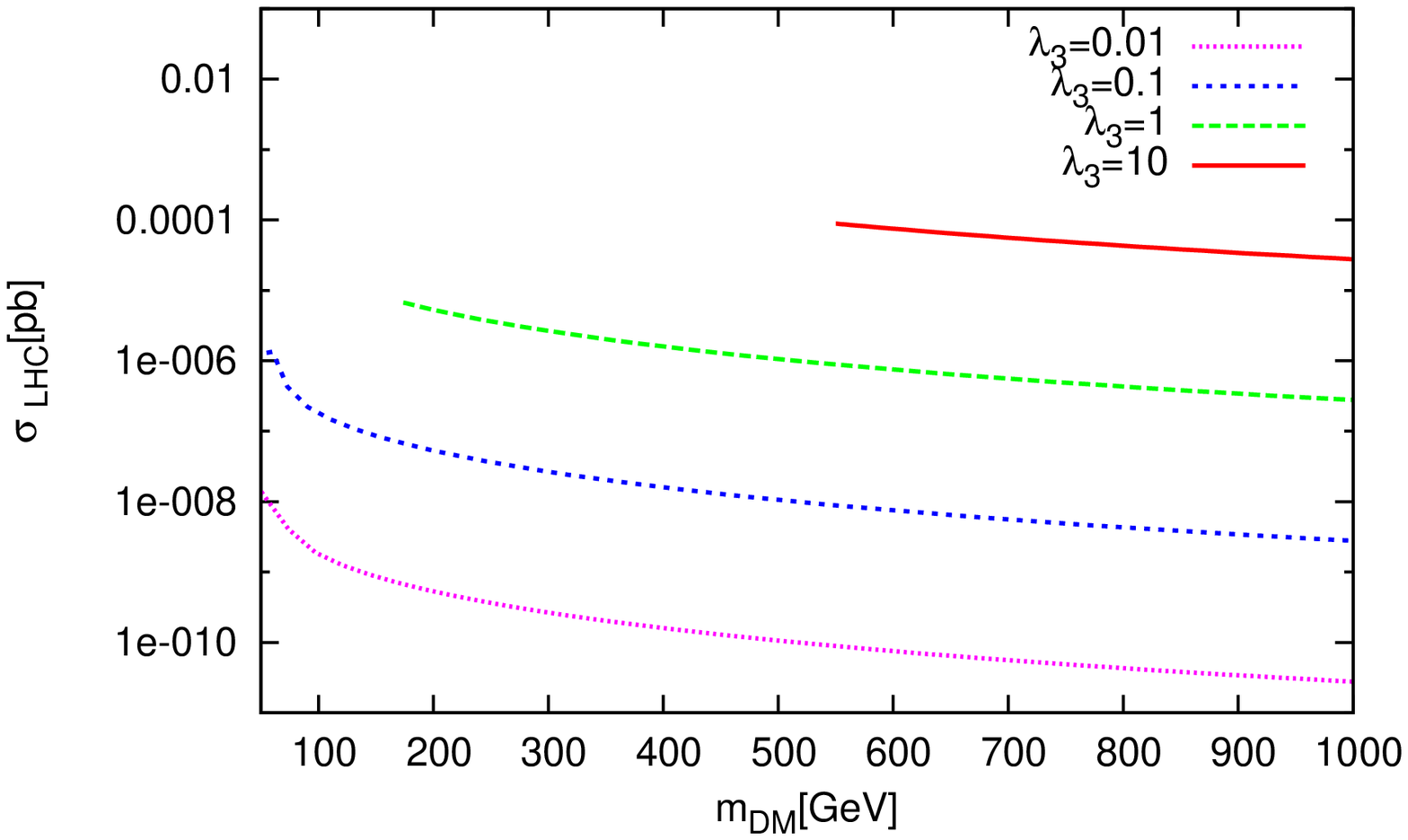,width=6.5cm}\hspace{0cm}\epsfig{figure=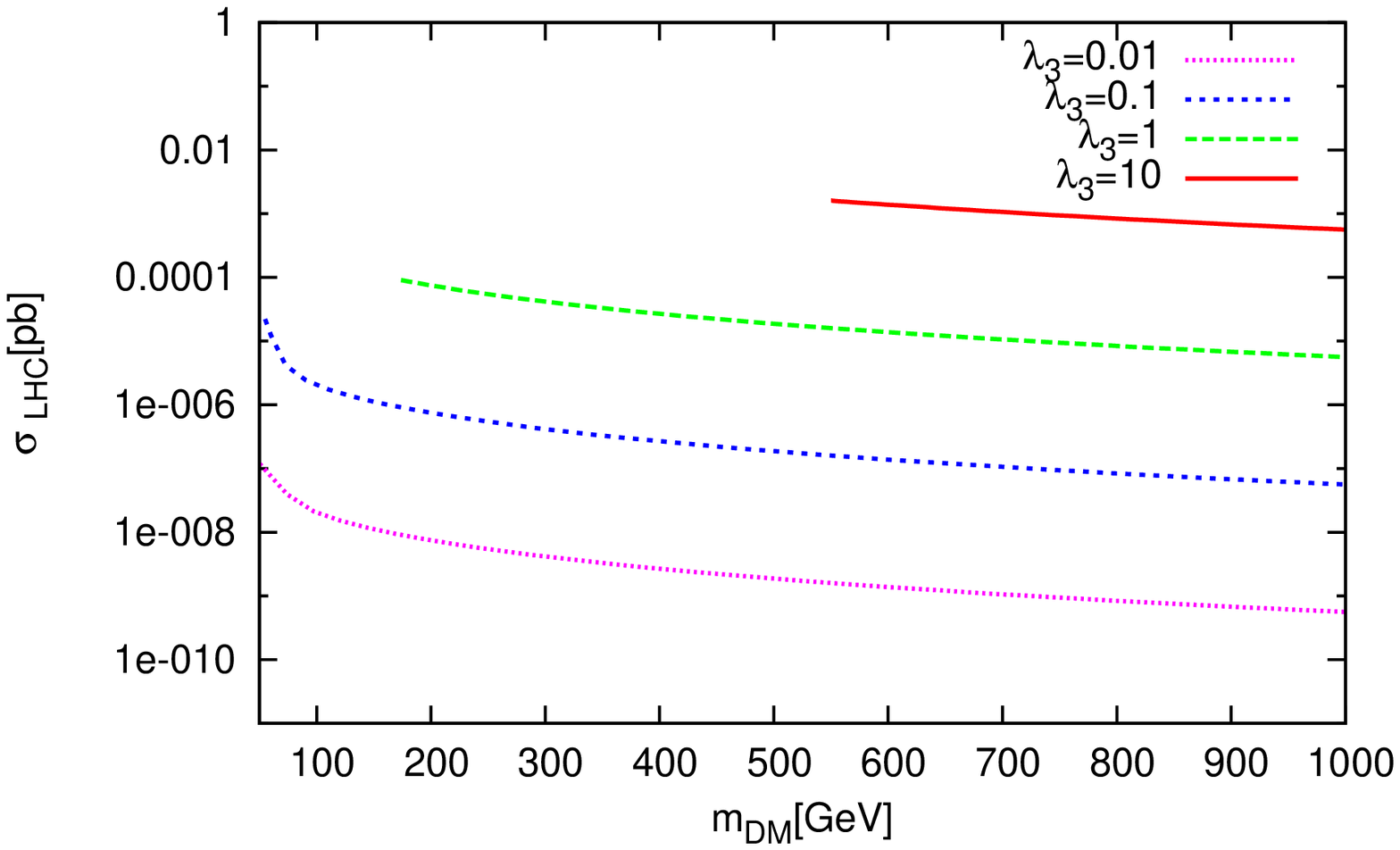,width=6.5cm}}
\centerline{\vspace{-2cm}\hspace{0.5cm}(a)\hspace{6cm}(b)}
\centerline{\vspace{0.8cm}}
\end{center}
\caption{The mono-Higgs production cross section as a function of
the DM mass a) at the LHC for the center-of-mass energy of $14~\rm TeV$ b) at the VLHC for the center-of-mass energy of $100~\rm TeV$.}\label{monoHiggs}
\end{figure}

In Fig.~\ref{monoHiggs}, we display total cross section for production of $hT^0T^0$  versus $m_{T^0}$ at the LHC and the proposed high energy collider with $100~\rm TeV$ center-of-mass energy. We expect that $\sigma(pp\rightarrow hT^0T^0)$ is very small because it is proportional to Yukawa couplings of $u$ and $d$ quarks which have main contribution in PDFs. Notice that the minimum value for $m_{T^0}$ is $\sqrt{1/2\lambda_3 v^2}$ (see Eqe.~\ref{mDM}).
If in the coming run of the LHC, $300~fb^{-1}$ data is collected, we can not probe DM-Higgs boson coupling in all regions of parameters space. One of the ideas for searching  beyond SM in future colliders is  Very Large Hadron Collider(VLHC) with center-of-mass energy at $100~\rm TeV$ \cite{VLHC}. As it is shown in Fig.~\ref{monoHiggs}-b, the total cross section of mono-Higgs in context of ITM can be in $\rm pb$ order. This means with $1000~\rm fb^{-1}$ data, we will detect statistically significant number of such events. As a result of precise measurement, it will become  possible to probe DM signatures for $m_{DM}>m_h/2$ in VLHC.

\section{Annihilation of Dark Matter into monochromatic photons}
\begin{figure}
\begin{center}
\centerline{\hspace{1cm}\epsfig{figure=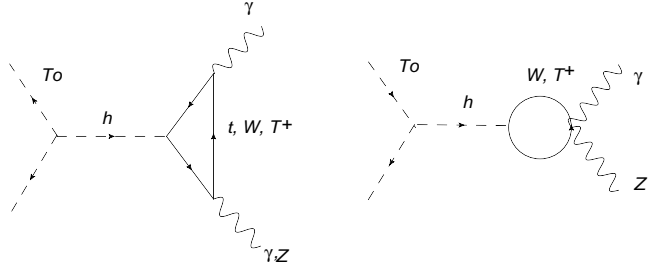,width=8.5cm}}
\centerline{\vspace{-1.5cm}}
\end{center}
\caption{The Feynman diagrams for annihilation of $T^0$ into $\gamma\gamma$ and $\gamma Z$} \label{anhi}
\end{figure}

In the following, we calculate possible annihilation of DM candidate in ITM into $2\gamma$ and $Z \gamma$. The Feynman diagrams for annihilation of $2T^0$ to $\gamma \gamma$ and $Z \gamma$ are shown in Fig.~\ref{anhi}. As it is seen, $T^{\pm}$ can contribute to these processes. The amplitude for $2T^0\rightarrow2\gamma$ can be written down as follows:
\begin{eqnarray}
 i{\cal{|M|}}_{2T^0\rightarrow2\gamma}=\frac{iv_0\lambda_3}{s-m^2_h-i m_h\Gamma_h}{\cal{M}}_{h\rightarrow2\gamma},
\label{amp1}
\end{eqnarray}
where $\sqrt{s}$ is the center-of-mass energy. The cross section is given by:
\begin{eqnarray}
\sigma\upsilon&=&\frac{1}{4\pi s}{\cal{|M|}}^2_{2T^0\rightarrow2\gamma} =\frac{1}{(s-m^2_h)^2+m_h^2\Gamma^2_h}\times\frac{\alpha^2g^2v^2_0\lambda^2_3s}{512\pi^3M^2_W}|\frac{4}{3}{\cal{A}}_{1/2}(x_i)\nonumber\\&+&{\cal{A}}_{1}(x_i)+2v_0\lambda_3\frac{ M_W}{gM^2_{T^{\pm}}}{\cal{A}}_{0}(x_i)|^2,
\label{cross1}
\end{eqnarray}
where ${\cal{A}}_{i}$  form factors have been given in Eqe.~\ref{function1} appendix.
Similar annihilation to  $Z\gamma$ is possible in context of ITM. The amplitude for $2T^0\rightarrow Z\gamma$ can be expressed as follows:
\begin{eqnarray}
 i{\cal{|M|}}_{2T^0\rightarrow Z\gamma}=\frac{iv_0\lambda_3}{s-m^2_h-i m_h\Gamma_h}{\cal{M}}_{h\rightarrow Z\gamma}.
\label{amp2}
\end{eqnarray}
The annihilation cross section into $Z\gamma$ is given by:
\begin{eqnarray}
\sigma\upsilon&=&\frac{1}{8\pi s}(\sqrt{1-\frac{m^2_Z}{s}}){\cal{|M|}}^2_{2T^0\rightarrow Z\gamma} =\frac{\alpha^2g^2v^2_0\lambda^2_3 }{64\pi^3 c^2_W}(1-\frac{m^2_Z}{s})^{5/2}|{\cal{A}}_{t}(x_i,y_i)\nonumber \\&+&{\cal{A}}_{W}(x_i,y_i)+{\cal{A}}_{T^+}(x_i,y_i)|^2\times\frac{1}{(s-m^2_h)^2+m_h^2\Gamma^2_h},
\label{cross2}
\end{eqnarray}
where ${\cal{A}}_{t}(x_i,y_i)$, ${\cal{A}}_{W}(x_i,y_i)$ and ${\cal{A}}_{T^+}(x_i,y_i)$  are given in Eqs.~\ref{formfactor} of the appendix.
FermiLAT collaboration has measured flux for diffuse gamma-ray background and  gamma-ray spectral lines from $7$ to $300~\rm GeV$  obtained from 3.7 years data. The cross section upper limit and decay lifetime lower limits for DM particle that produce gamma-ray lines or contribute to the diffuse spectrum have been given in\cite{FermiLat} and \cite{FermiLat1}. Here, we compute thermal average cross section of annihilation and use these data to put constraint on ITM parameter space. The thermal average cross section is expressed by\cite{thermal average}:
\begin{eqnarray}
\langle\sigma\upsilon\rangle=\frac{1}{8m^4_{DM}T_FK^2_2(m_{DM}/T_F)}\int^\infty_{4m^2_{DM}}ds\sigma(s)(s-4m^2_{DM})\sqrt{s}K_1(\frac{\sqrt{s}}{T_F})
\label{thermal}
\end{eqnarray}
where $K_n(x)$ is the modified Bessel function and inverse scale freeze-out temperature $x_F$ is estimated by following equation:
\begin{eqnarray}
x_F&=&\ln(\frac{0.382c m_{DM}M_{pl}g_{DM}}{\sqrt{g_*x_F}}\langle\sigma\upsilon\rangle)
\label{xf}
\end{eqnarray}
where $x_F=m_{DM}/T_F$, $c=\sqrt{2}-1$ and $g_*=91.5$ \cite{godebole}.
As it is mentioned, in ITM the DM candidate can annihilate into $2\gamma$ and $Z \gamma$ at the loop level. If DM annihilate directly into two photon
or a photon and another particle (X), the photons energy are respectively:
 \begin{eqnarray}
E_{\gamma}=m_{DM}  ~~ {\rm and} ~~\ E_{\gamma}=m_{DM}(1-\frac{m^2_X}{4m^2_{DM}}).
\label{Ephoton}
\end{eqnarray}

\begin{figure}
\begin{center}
\centerline{\hspace{-0.5cm}\epsfig{figure=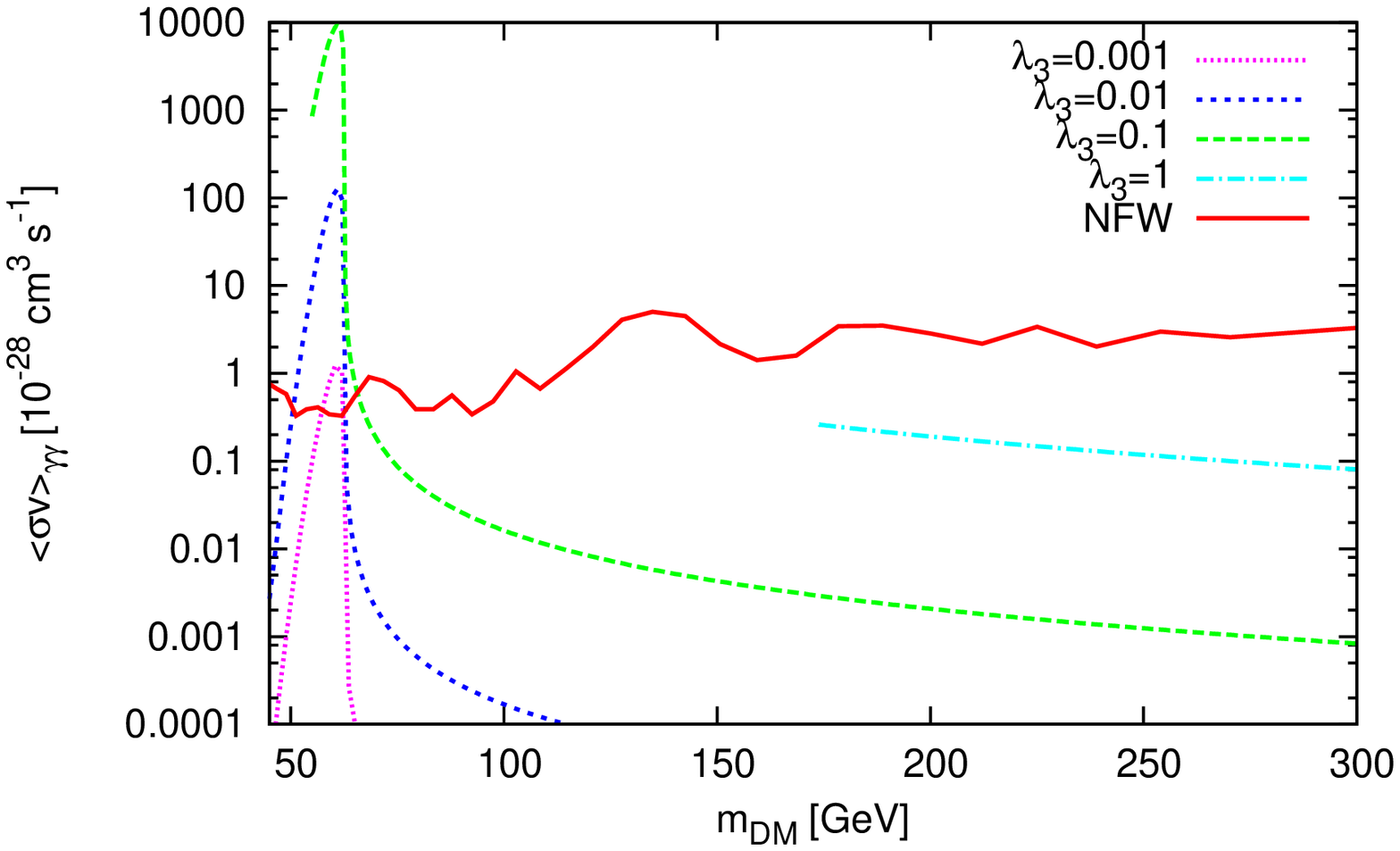,width=6.5cm}\hspace{0cm}\epsfig{figure=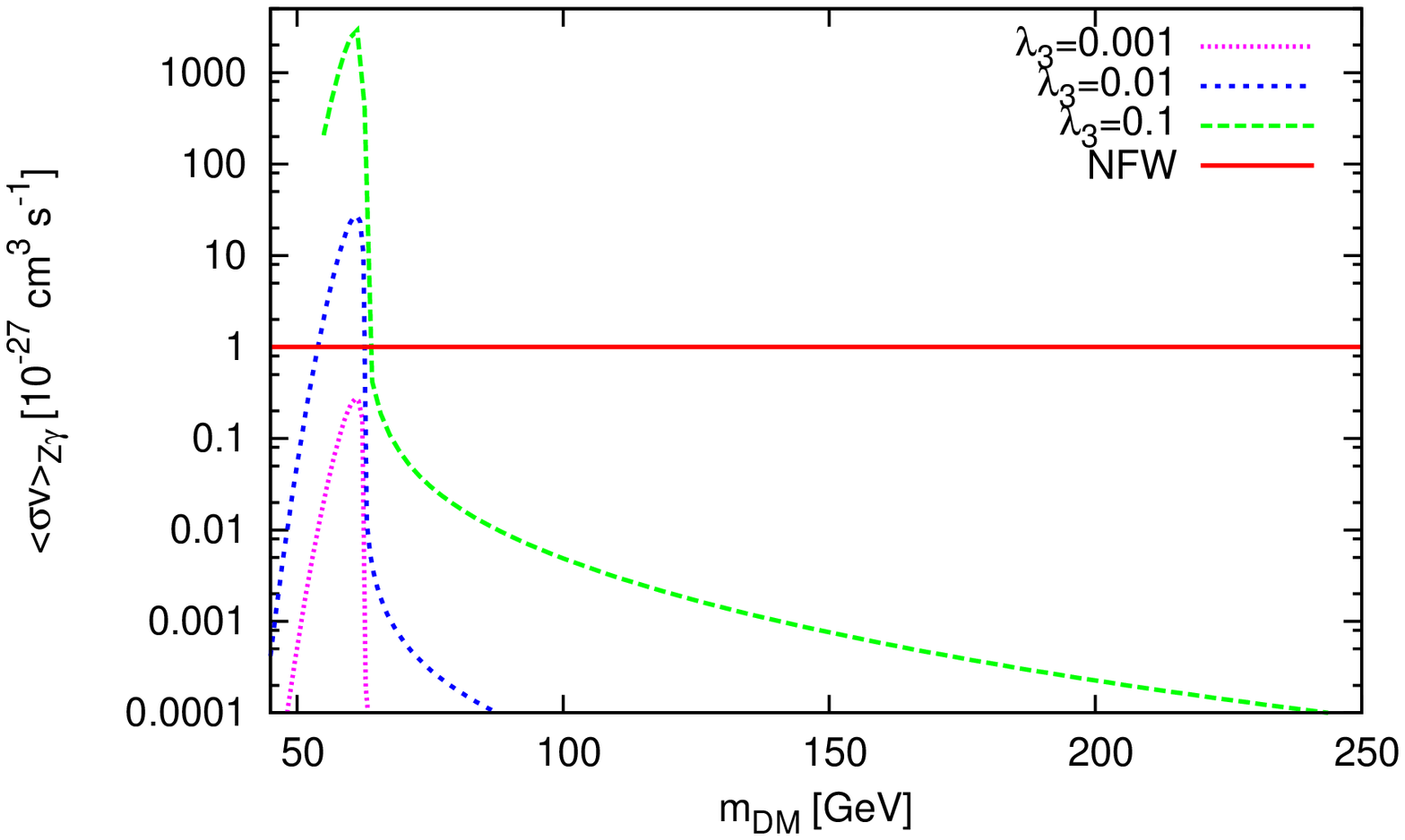,width=6.5cm}}
\centerline{\vspace{-2cm}\hspace{0.5cm}(a)\hspace{6cm}(b)}
\centerline{\vspace{0.8cm}}
\end{center}
\caption{The thermal average annihilation cross-section of $T^0$ (DM) to (a) $\gamma\gamma$ and (b)$Z\gamma$ as a function of
the DM mass for several values of $\lambda_3$. The solid red lines shows the upper limits on annihilation cross-section which have borrowed from \cite{FermiLat}-\cite{FermiLat1}.}\label{sigmav}
\end{figure}

Now we consider the FermiLAT upper limit on annihilation cross-section for NFW, Einasto and Isothermal DM profiles in the Milky Way which have been presented in \cite{FermiLat} for $\gamma\gamma$ and in \cite{FermiLat1} for $Z\gamma$ annihilation and study the constraints on the parameter space of ITM arising from these data. In Fig.~\ref{sigmav}-a(b) we plot the thermal average cross section for annihilation of DM to $\gamma\gamma$($Z\gamma$) as a function of
the DM mass for several values of $\lambda_3$. In this figure, for process $T^0T^0\rightarrow \gamma\gamma$($T^0T^0\rightarrow Z\gamma$), we suppose $E_{\gamma}=m_{DM}$($E_{\gamma}=m_{DM}[1-m^2_Z/4m^2_{DM}$]). The solid red lines shows the upper limits on annihilation cross-section for NFW density profile in the Milky Way which have borrowed from \cite{FermiLat}-\cite{FermiLat1}. For the case $2T^0\rightarrow Z\gamma$, red line shows minimum value for upper limit on annihilation in $m_{DM}=344~\rm GeV$ and $\langle\sigma\upsilon\rangle=1.2\times10^{-27}~\rm cm^{3}s^{-1}$. Note that from \cite{FermiLat1}, for  $m_{DM}<235~\rm GeV$, upper limits on $\langle\sigma\upsilon\rangle$ is not accessible.

\begin{figure}
\begin{center}
\centerline{\hspace{-0.5cm}\epsfig{figure=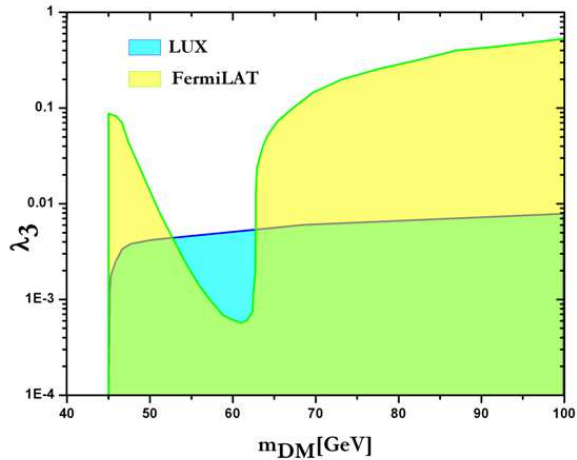,width=8.5cm}}
\centerline{\vspace{-1.2cm}}
\end{center}
\caption{Shaded areas depict ranges of parameter space in mass of DM and $\lambda_3$ coupling plane which are consistent with
 upper limit on $\sigma_{\rm FermiLAT}$ with $95\%$ C.L (indirect detection) and $\sigma_{\rm LUX}$ with $90\%$ C.L (direct detection)}\label{scaterindirect}
\end{figure}

 As it can be seen, for $m_{DM}>63~\rm GeV$, $\langle\sigma\upsilon\rangle$ is very smaller than upper limits on annihilation cross-section and it would not constrain ITM parameters space. However, for $m_{DM}<63~\rm GeV$ (near to the pole of Higgs propagator at  $m_{DM}=m_h/2$), the cross section increases (see Eqe.~\ref{cross1}) and will be larger than upper limit. To achieve most conservative limit on $\lambda_3$, we consider the minimum upper limit on $\sigma_{\rm FermiLAT}=0.33\times 10^{-28}$ with $95\%$ C.L for NFWc profile \cite{FermiLat1}. In Fig.~\ref{scaterindirect}, We have shown that allowed region on DM mass and $\lambda_3$ coupling plane which are consistent with this limit and LUX bound. It is notable that FermiLAT constraint is stronger than direct detection limit in region $52<m_{DM}<63$. Since for $T^0T^0\rightarrow Z\gamma$, energy of photon is equal to $m_{DM}[1-m^2_Z/4m^2_{DM}]$, upper limit on cross section is weaker than the one's from $T^0T^0\rightarrow \gamma\gamma$ (with similar DM mass) and as a result constraints on $\lambda_3$ are much weaker.

\section{Conclusions}

In this paper, we have investigated an extension of SM which includes a $SU(2)_L$ triplet scalar with hypercharge $Y=0,2$. This model provide suitable candidate for DM, because the lightest component of triplet field is neutral and for the $m_{DM}<7~\rm TeV$, conditions of relic abundance are satisfied. We focus on  parameter space  which is allowed by PLANCK data and study collider phenomenology of inert triplet scalar DM at the LHC.

We have shown that the effect of ITM on invisible Higgs decay for low mass DM ($m_{DM}<63~\rm GeV$) can be as large as constraints from LUX direct detection  experiment ( see Fig.~\ref{scater}-a). We also have shown that contribution of ITM DM to $R_{\gamma\gamma}$  can be comparable with direct detection and invisible higgs decay.

We also study effect of ITM DM on $R_{Z\gamma}$ and mono-Higgs production at the LHC.  The sensitivity of experimental measurement for $R_{Z\gamma}$ is far from SM prediction. We have shown in Fig.~\ref{RZgamma}, this effect is in same order of magnitude of SM prediction (smaller than 1). Also for mono-Higgs production ITM effect is very small.  For these reasons, at present time these observables can not constrain ITM parameter space.

Eventually, We calculate the annihilation cross section of DM candidate into $2\gamma$ and $Z\gamma$. The minimum upper limit on annihilation cross-section from FermiLAT have been employed to constraint parameters space of ITM. We also compared our results with direct detection DM constraints and showed for $52<m_{DM}<63~\rm GeV$,  FermiLAT constraint is stronger than direct detection constraint for low mass DM.

\section{Acknowledgement}
The authors would like to thank M. Torabian and  M. Tavakoli for the useful discussions.
We also acknowledge to S. Paktinat for careful reading of the
manuscript and the useful remarks.

\section{Appendix}
In this appendix, we summarize the formulae of form factors which contribute to Higgs decay rate $\gamma\gamma$, $Z\gamma$  and annihilation cross section of $T^0$ to $\gamma\gamma$, $Z\gamma$.
The form factors ${\cal{A}}_{i}(x_j)$ which arise from triangle diagram in  $h\rightarrow\gamma\gamma$ can be expressed by:
\begin{eqnarray}
{\cal{A}}_{0}(x_i)&=&-[x_i+f(x_i)]x^{-2}_i
\nonumber\\{\cal{A}}_{1/2}(x_i)&=&2[x_i+(x_i-1)f(x_i)]x^{-2}_i
\nonumber\\{\cal{A}}_{1}(x_i)&=&-[3x_i+2x^2_i+3(2x_i-1)f(x_i)]x^{-2}_i,
\label{function1}
\end{eqnarray}
where $x_i=\frac{m^2_h}{4m^2_i}$ and $f(x_i)$ is:
\begin{eqnarray}
  f(x) =\left\{
  \begin{array}{ll}  \displaystyle
    (\arcsin \sqrt{x})^2 & x\leq 1 \\
    \displaystyle -\frac{1}{4}\left[ \log\frac{1+\sqrt{1-x^{-1}}}
    {1-\sqrt{1-x^{-1}}}-i\pi \right]^2 \hspace{0.5cm} & x>1
  \end{array} \right.
\label{f}
\end{eqnarray}

Let us consider $Z\gamma$ decay. The $Z\gamma$ form factors ${\cal{A}}_{t}(x_i,y_i)$, ${\cal{A}}_{W}(x_i,y_i)$ and ${\cal{A}}_{T^+}(x_i,y_i)$ are given by:
\begin{eqnarray}
{\cal{A}}_{T^+}(x_{T^+},y_{T^+})&=&\frac{2(2c^2_W-1)}{c^2_W}\frac{gM_Wv_0\lambda_3}{M^2_{T^+}}I_1(x_{T^+},y_{T^+})
\nonumber\\{\cal{A}}_{t}(x_{t},y_{t})&=&\frac{2-(16/3)s^2_W}{s_W c_W}[I_1(x_{t},y_{t})-I_2(x_{t},y_{t})]
\nonumber\\{\cal{A}}_{W}(x_W,y_W)&=&\cot\theta_W\{4(3-\tan\theta^2_W)I_2(x_W,y_W)\nonumber \\&+&[(1+2x_W)\tan\theta^2_W-(5+2x_W)]I_1(x_W,y_W)\},
\label{formfactor}
\end{eqnarray}
the arguments $x_i$ and $y_i$ are $x_i=\frac{m^2_h}{4m^2_i}$ and $y_i=\frac{m^2_Z}{4m^2_i}$.  Functions $I_{i}(x,y)$ are defined by:
\begin{eqnarray}
I_{1}(x,y)&=&\frac{-1}{2(x-y)}+\frac{1}{2(x-y)^2}[f(x)-f(y)]+\frac{y}{2(x-y)^2}[g(x)-g(y)]
\nonumber\\I_{2}(x,y)&=&\frac{1}{2(x-y)}[f(x)-f(y)],
\label{I}
\end{eqnarray}
where function $f(x)$ is defined in Eqe.~\ref{f} and $g(x)$ is:
\begin{eqnarray}
 g(x) =\left\{
  \begin{array}{ll}  \displaystyle
    \sqrt{x^{-1}-1}\arcsin \sqrt{x} & x\leq 1 \\
    \displaystyle \frac{1}{2}\sqrt{1-x^{-1}}\left[ \log\frac{1+\sqrt{1-x^{-1}}}
    {1-\sqrt{1-x^{-1}}}-i\pi \right]. \hspace{0.5cm} & x>1
  \end{array} \right.
\end{eqnarray}

\end{document}